\documentclass[twocolumn,pre,amsmath,amssymb]{revtex4}

\usepackage{graphicx}
\begin{document}                  

\title{X-ray Studies of the Phases and Phase Transitions of Liquid Crystals}

\author{P.S.~Clegg}
\affiliation{School of Physics, University of Edinburgh, Mayfield Road,
Edinburgh EH9 3JZ, UK}

\begin{abstract}
This is a short review of recent x-ray diffraction studies of the phases
and phase
transitions of thermotropic liquid crystals. The areas covered are
twist-grain-boundary phases,
antiferroelectric phases studied with resonant x-ray
diffraction, and smectic phases within gel structures. 
In all areas x-ray diffraction has played a key role. Nonetheless open
questions remain: the nature of the smectic-C variant of the
twist-grain-boundary phase, the origin of antiferroelectric phases, and
whether novel Bragg glass states exist for smectic-A gel samples.
\end{abstract}

\maketitle                        

\section{Introduction}

It was first suggested that smectic liquid crystals are ordered in layers
based on observations of large scale defect structures in the 1920's 
as described by
de Gennes and Prost (1993). However the novel structure of the 
smectic-A (SmA) phase
(Als-Nielsen \textit{et al.}, 1980) and the character of the nematic (N)
to SmA transition (Davidov \textit{et al.}, 1979) 
were established using x-ray diffraction
complemented by other techniques. X-ray diffraction also played a decisive
role in elucidating the properties of the SmA - smectic-C (SmC) transition
(Safinya \textit{et al.}, 1980) where the molecules acquire a tilt with
respect to the layer normal. Similarly, it was an important technique
for the SmA - chiral SmC (SmC$^*$) transition (Shashidhar \textit{et
al.}, 1988).
Major reviews of x-ray studies of liquid crystals have been
carried out by Pershan (1988) and Seddon (1998). The latter gives a
comprehensive list
of reviews of structural studies of liquid crystals.
A recent review by Kumar (2001) focuses on experimental techniques.
The selection of topics included here is intended to be
representative rather than complete.
The topics concern the perturbation of
SmA and SmC phases by twist, dipole moments and local deformations.
Novel phases and phase transition behaviour results.

Section II focuses on twist-grain-boundary (TGB) phases especially
those with SmC order. The TGB phase was proposed by Renn and Lubensky
(1988) and discovered by Goodby \textit{et al.} (1989a, 1989b). These
are smectic phases where arrays of defects form part of the ordered
structure. The layer order can have SmA, SmC or SmC$^*$
structure. The defects are a means of incorporating the twist of chiral
molecules into a layered phase. More recently large scale
superstructures have been observed in the plane perpendicular to the
helical axis (Pramod \textit{et al.}, 1997; Ribeiro \textit{et al.},
1999). The nature of the SmC and SmC$^*$ TGB
structures and their relationship to larger scale superstructures are still
open issues (Brunet \textit{et al.}, 2002).

Section III concerns the family of antiferroelectric phases.
Antiferroelectricity in stacks of two dimensional liquids was not
predicted before its discovery. Initial experimental signatures were not
immediately associated with this phenomenon 
(Hiji \textit{et al.}, 1988; Furukawa \textit{et al.}, 1988). 
Subsequently a series of
antiferroelectric phases were discovered (Fukuda \textit{et al.}, 1994).
These include both commensurate and incommensurate stacking sequences.
Resonant x-ray scattering has enabled these sequences to be determined
(Mach \textit{et al.}, 1998).
The structures occur for materials with large tilts, large dipole
moments and no cholesteric phase. The fundamental question of the 
origin of the phases is still not resolved
(Lagerwall \textit{et al.}, 2003).

Section IV covers studies of the phase transitions of liquid
crystals confined in gels. Early experiments were carried out for the
N-SmA transition in rigid aerogels (Clark \textit{et al.}, 1993).
Enhanced
anomalous elastic effects and a smectic Bragg glass state were later
predicted (Radzihovsky \& Toner, 1999). Studies in rigid
aerogels (Bellini \textit{et al.}, 2001) and soft aerosil gels 
(Park \textit{et al.}, 2002) have been made to test this model. Smectic
Bragg glass signatures from aerogel samples (Bellini \textit{et al.},
2001) have not reappeared with aerosil samples
under conditions where they should be more
evident (Leheny \textit{et al.}, 2003; Clegg \textit{et al.}, 2003a).

\section{Twist-grain-boundary phases}

De Gennes (1972) identified an
analogy between the Meisner state of a superconductor and the
smectic state of a liquid crystal.
In the Meisner state, magnetic field is expelled from
the solid; equivalently twist is expelled from the smectic state.
Abrikosov (1957) predicted a vortex solid state in type II superconductors
in an applied magnetic field. De Gennes (1972) hinted at and
Renn and Lubensky (1988) described the analogous
phase for a liquid crystal.
In the Abrikosov vortex lattice, magnetic
field penetrates the sample at defects (vortices) which themselves
order in a regular array. On heating, the state returns to a normal metal
via a vortex liquid phase. Twist is introduced into a smectic when the
molecules have chirality and a single enantiomer is present. This
contributes a term
\begin{equation}
f_{Ch} = -h \mathbf{\hat{n}.\nabla} \times \mathbf{\hat{n}}
\end{equation}

\noindent to the Frank free energy density (which describes the energy
cost of distortions to the orientational order). Here, $h$ is the strength of
the twist and $\mathbf{\hat{n}}$ is the nematic director.
Renn and
Lubensky (1988) detailed a
twist-grain-boundary state with blocks of well ordered smectic separated
from one another by walls of regularly spaced dislocations. The layer
normal is offset by an angle $\Delta$
from one block to the next with an axis parallel to the
planes. The angle is given by $d = 2\ell_d \sin(\Delta / 2)$,
where $d$ is the thickness of the smectic layers and $\ell_d$ is the
separation between screw dislocation lines.
The structure thus has a helical axis and twist has
been incorporated into the phase.
Chiral liquid crystals often
have a cholesteric phase at high temperatures; a full pursuit of the
analogy would suggest that a twist line liquid phase, N$^*_L$, 
should occur between
the cholesteric phase and the twist-grain-boundary phase in parallel to
the vortex liquid phase. Kamien and Lubensky (1993) have discussed the
structure of the N$^*_L$ phase, concluding that the screw dislocation
lines should themselves have cholesteric order.

\begin{figure}[h]
\includegraphics[scale=0.8]{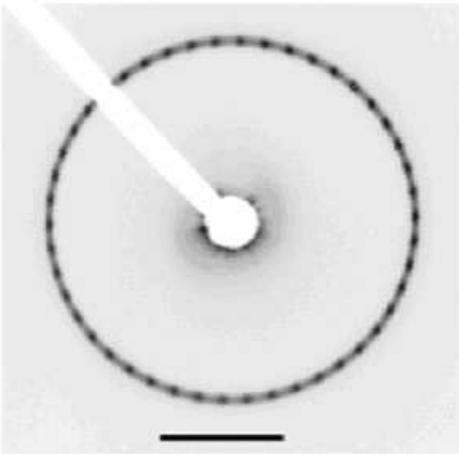}
\caption{The modulated x-ray intensity due to the TGBA phase. In this
case the ring has 64 spots. (Reproduced with 
permission from Navailles \textit{et al.}, 1998.)}
\end{figure}

\subsection{Twist-grain-boundary SmA and N$^*_L$ phases}

The TGB phase was discovered by Goodby \textit{et al} (1989a, 1989b)
following the theoretical impetus.
For the x-ray studies on oriented samples, which followed, very long
correlation lengths were required to see many
ordered blocks of twist-grain-boundary phase. Sample holders tens of
microns thick with rubbed surfaces were used to create the orientational
order.
Detailed studies of the TGB SmA (TGBA) phase 
and the TGBA to N$^*_L$ transition
were carried out by Navailles \textit{et al.} (1998).
The TGBA phase was found to be
made up of a series of blocks of SmA order each
with a layer normal which is perpendicular to the helical axis (as
implied by Fig. 1). The
layer normal rotates in discrete steps from one block to the next at a
grain boundary.
It was shown that the size of SmA blocks, $\ell_b$, in
the TGBA phase was approximately equal to the distance between the screw
dislocations, $\ell_d$, 
in the grain boundaries as predicted (Renn \& Lubensky,
1988). 

Measurements using ac-calorimetry revealed a large $\Delta C_p$ peak
(Chan \& Garland, 1995) consistent with the
cholesteric phase evolving into a twisted line liquid phase N$^*_L$. The
warming of the TGBA phase into the N$^*_L$ phase is the analogue of the
Abrikosov vortex lattice melting and here was shown to be a first-order
transition.
The transition from the vortex solid phase to the vortex liquid
phase TGBA to N$^*_L$ was found to be marked by the helical twist angle
becoming incommensurate (Navailles \textit{et al.}, 1998) 
and a steady decrease in the SmA order parameter
(here the parallel correlation length, $\xi_{\|}$, is reported to
remain unchanged).
Further studies of the transition from the TGBA to N$^*_L$ phase have
been made by Ybert and coworkers (Ybert \textit{et al.}, 2003). 
They found the temperature dependence
of the smectic correlation length consistent with the melting via the
N$^*_L$ phase. The different result to that reported above is likely due
to an increase in the wave-vector resolution for the second study.

\subsection{Twist-grain-boundary SmC and SmC$^*$ phases}

Renn and Lubensky (1991)
predicted a SmC variant of the twist-grain-boundary
phase where the molecules acquire a tilt angle in the plane
perpendicular to the helical axis. Further complications can occur.
The SmC phase is modified by chiral molecules to give the SmC$^*$ phase.
The azimuthal angle associated with the tilt precesses from one plane to
the next. This creates a helix with a pitch $\sim1$ $\mu$m. With chiral
molecules there is
the possibility of a TGB phase with 
twist along two orthogonal axes as predicted
by Renn (1992) for the TGB SmC$^*$ (TGBC$^*$) phase.
Here the SmC slabs are grouped in larger
clusters called helislabs with a shared helical axis for SmC$^*$ order.

The TGB SmC (TGBC) phase was discovered by Nguyen \textit{et al.} (1992).
Historically, x-ray measurements on oriented samples 
were made on
TGBC phase before the TGBA phase. On
cooling from the N$^*$ phase into the TGBC phase the continuous ring of
scattering from short-range smectic correlations is seen to become
strongly modulated in intensity (Navailles \textit{et al.}, 1993). The
period of the modulation  varies discontinuously on cooling. Each peak
on the ring is due to scattering from one set of blocks of smectic
order. The orientation of these blocks jumps at the grain boundaries by
angles commensurate with the pitch of the helix.
Both the pitch of the helix and the
size of the blocks may vary with temperature.

It was predicted by Renn and Lubensky (1991) that
the layer normal would be perpendicular to the axis of the helix.
Experiments
showed that this is not the case (Navailles \textit{et al.}, 1995), 
at least not for this variant of the TGBC phase. The TGBC phase studied
in Bordeaux and described above is designated B-TGBC.
The situation is shown in Fig.2 with the layer normal at angle
$\omega_L$ to the direction of the molecules (which are orthogonal to the
helical axis). The tilt angle, $\omega_L$, grows from zero with a power
law on entering the TGBC phase.
This observation implies that the
grain boundaries are more complicated than a grid of screw dislocations.
Based on these results a model was created in which it was demonstrated
that in the TGBC structure described by Renn (1992) 
the director orientation can barely vary
across a SmC block (Dozov, 1995). This limits the amount of twist that can be
accommodated in this defect structure. Dozov (1995) proposed a structure with
a component of the layer normal in the TGB axis direction and with
melted grain boundaries. With this structure the director can then
precess across a block while maintaining a constant angle to the
planes.

A further anomaly with the B-TGBC structure is that it has only been
observed as a commensurate structure (the ratio of the helical pitch
$\Lambda = 2\pi \ell_b / \Delta$ 
to the block thickness $\ell_b$ is always an integer). 
With an incommensurate TGBC phase
the plane spacing would be observed to decrease without the discrete
spots developing. The fact that the structure is always commensurate
implies that there are surprisingly long-range interactions between the
SmC blocks or between the defects in the boundaries (Galerne, 2000).

\begin{figure}[h]
\includegraphics[scale=0.6]{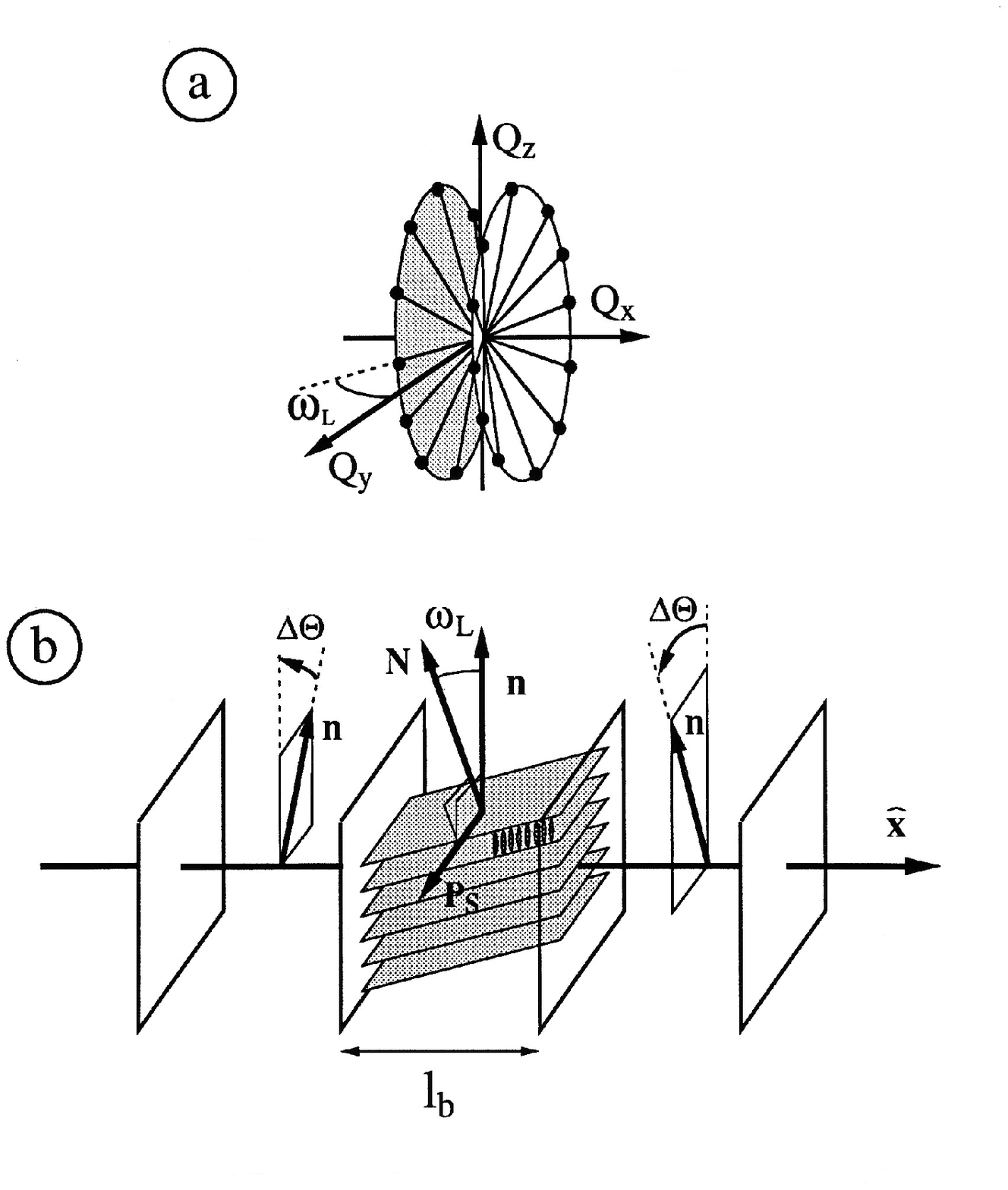}
\caption{The structure of the TGBC phase as seen in (a) reciprocal space and
(b) direct space. The smectic layers are tilted by angle $\omega_L$;
the director field and layer normal for the $i^{th}$ slab are
$n_i$ and $N_i$ respectively (Adapted from Petit \textit{et al.}, 1996).}
\end{figure}

Subsequently a range of other TGBC and TGBC$^*$ phases were
discovered. The most prominent feature with the new phases 
is that order is
observed on $\mu$m length scales 
in the plane perpendicular to the TGB axis. The first to be
found was reported by Pramod \textit{et al.} (1997).
The TGBC$^*$ phase was
observed in a binary mixture and exhibited a square grid pattern of in-plane
undulations; it is designated U-TGBC$^*$. Subsequently x-ray
studies were made on a single component TGBC material which also
exhibited a square superstructure in-plane (Ribeiro \textit{et al.},
1999). The spots from the SmC
blocks were substantially broadened in the TGB axis direction suggesting
that the direction of the plane normal was varying within a block.
It has not been determined whether the structure is commensurate.
This
phase was characterized by a team in Strasbourg and hence is denoted the
S-TGBC phase. A brief account has also been published (Brunet \textit{et
al.}, 2002) of a TGBC phase in
which the normal to the plane is perpendicular to the TGB axis as
predicted by Renn and Lubensky (1991). 
In this case six spots are observed from the SmC
blocks and there is a simultaneous hexagonal grid pattern.

Galerne (2000) and subsequently Brunet \textit{et al.}
(2002) have attempted to unify
these results by extending Renn's model of the TGBC$^*$ phase (Renn,
1992). 
Galerne
has shown that the helislab structure remains possible when the director
is predominantly perpendicular to the TGB axis. 
For SmC$^*$ blocks the
component of the director in the smectic plane, $\mathbf{c}$, precesses
around the layer normal, $\mathbf{N}$. This can occur if there is a
modulation of $\mathbf{n}$ with wave vector $\mathbf{q}$ in the yz plane
and a simultaneous modulation of $\mathbf{N}$ with the same wave vector
but in the xz plane. The second modulation must have a phase shift of
$\pi / 2$ from the first. This construction gives SmC$^*$ order but
requires that there is a compression and dilation of the layers. This
could be possible for small smectic blocks and indeed is consistent with
the peak broadening observed for the S-TGBC phase.
It is shown that angular lock-in may occur due to elastic
interactions between disclination lines. If so the helislabs will be 
commensurate and
the disclination lines from different boundaries will overlay each
other. Disclination lines scatter light and so will be visible using
microscopy. For large commensurate angles
superstructures will be observed. This is consistent with some of the known
behaviour of TGBC$^*$ materials.

Brunet \textit{et al.} (2002)
introduced a perspective on TGBC and TGBC$^*$ structures based on
experience with SmC$^*$ materials sandwiched between glass plates. The
model presented is somewhat similar to those of Renn (1992) and Galerne (2000) 
except that the
helislabs are the same thickness as the smectic blocks. The blocks are
polarization-splayed smectic slabs with boundaries that have a lattice
of unwinding lines on either side. Overlapping unwinding lines give the
observed superstructures.
The experimental results might be consistent with the models of Galerne
(2000) and Brunet \textit{et al.} (2002) with
different size helislabs for the different structures observed.

\section{Resonant x-ray studies of antiferroelectric phases}

Chiral molecules reduce the symmetry of the SmC phase by facing, on average,
a particular in-plane direction (Meyer \textit{et al.}, 1975). 
For molecules with a transverse dipole
moment (which is the usual case) this gives the ferroelectric SmC$^*$ phase.
The structure has a slight
precession of the tilt giving a helix with a pitch $\sim1$ $\mu$m. An
antiferroelectric SmC$^*_A$ phase was discovered in the late 1980's
during attempts to synthesise new ferroelectric materials (Hiji
\textit{et al.}, 1988; Furukawa \textit{et al.}, 1988) as described
by Fukuda \textit{et al.} (1994).
The tilt and
transverse dipole moment directions alternate by close to $\pi$
from one plane to the next, the deviation being due to the long helical
pitch (Fig.3).
Further cooling from this phase 
leads directly to either a solid
crystal phase or a hexatic phase. 
Subsequently new phases have been discovered which are
stable both at higher temperature than the SmC$^*$ phase (SmC$^*_{\alpha}$)
and also in the temperature range between the SmC$^*$ and SmC$^*_A$
phases (Fukuda \textit{et al.}, 1994). In the latter range 
two phases
have been identified. One of the most powerful techniques for unraveling the
nature of these phases is resonant x-ray diffraction (Mach \textit{et
al.}, 1998).
It has been
possible to discriminate between plane ordering configurations for
stacked planes of liquids. X-ray reflectivity studies have also been
employed to study the stacking sequence of thin films (Fera \textit{et
al.}, 2001). Conventional x-ray diffraction studies of higher
order reflections have probed the interdigitation of smectic layers
(Takanishi \textit{et al.}, 1995).
The observed phases are now the subject of
theoretical research.

\subsection{Resonant x-ray diffraction}

X-ray scattering at energies close to an atomic absorption edge reveals
information about the environment of that atom. It is called resonant
scattering and it has been used for two decades to study solid crystals.
A key feature was demonstrated by Templeton and Templeton (1982) who 
showed that resonant scattering depends
on the polarization direction for anisotropic molecules.
The reason for this is that resonant scattering 
involves virtual transitions of
the inner shell electron to an intermediate state (Als-Nielsen \&
McMorrow, 2001). The resonant effect
probes the unoccupied states which are connected to the inner shell by
an electric dipole transition. Due to selection rules the possibility of
transitions depends on the relative orientation of the x-ray electric
field and the orbital.
Additional reflections can occur if the
anisotropy of the electron environment exhibits a new
periodicity.

Dmitrienko (1983) considered how resonant scattering would modify
the signature of glide planes and screw axes.
The form factor and
polarization conditions for the new scattering at previously forbidden
reflections were calculated. Levelut and Pansu (1999) 
extended the studies of
Dmitrienko:
they calculated the tensor structure factor for different
incommensurate helical phases and for specific plane sequence
commensurate structures. In practice $\mathbf{\sigma}$ polarized
radiation is usually used. The first-order
reflections are then predicted to be 
$\mathbf{\pi}$ polarized
whereas the second-order reflections are elliptically
polarized. In this case the balance between $\mathbf{\sigma}$ and
$\mathbf{\pi}$ depends on $\sin(\theta)$, where $2\theta$ is the
scattering angle, leading to predominantly
$\mathbf{\sigma}$ polarization. 
Levelut and Pansu (1999) note that models can
be discriminated using polarization analysis and that distortion angles
can be found from the relative intensities of the resonant reflections.

A synchrotron x-ray source is essential to obtain the continuous
spectrum of polarized radiation required for 
resonant diffraction. The low energies used for
the sulfur K-edge are strongly scattered by air. This implies that the
whole apparatus needs to be evacuated.

\begin{figure}[h]
\includegraphics[scale=0.32]{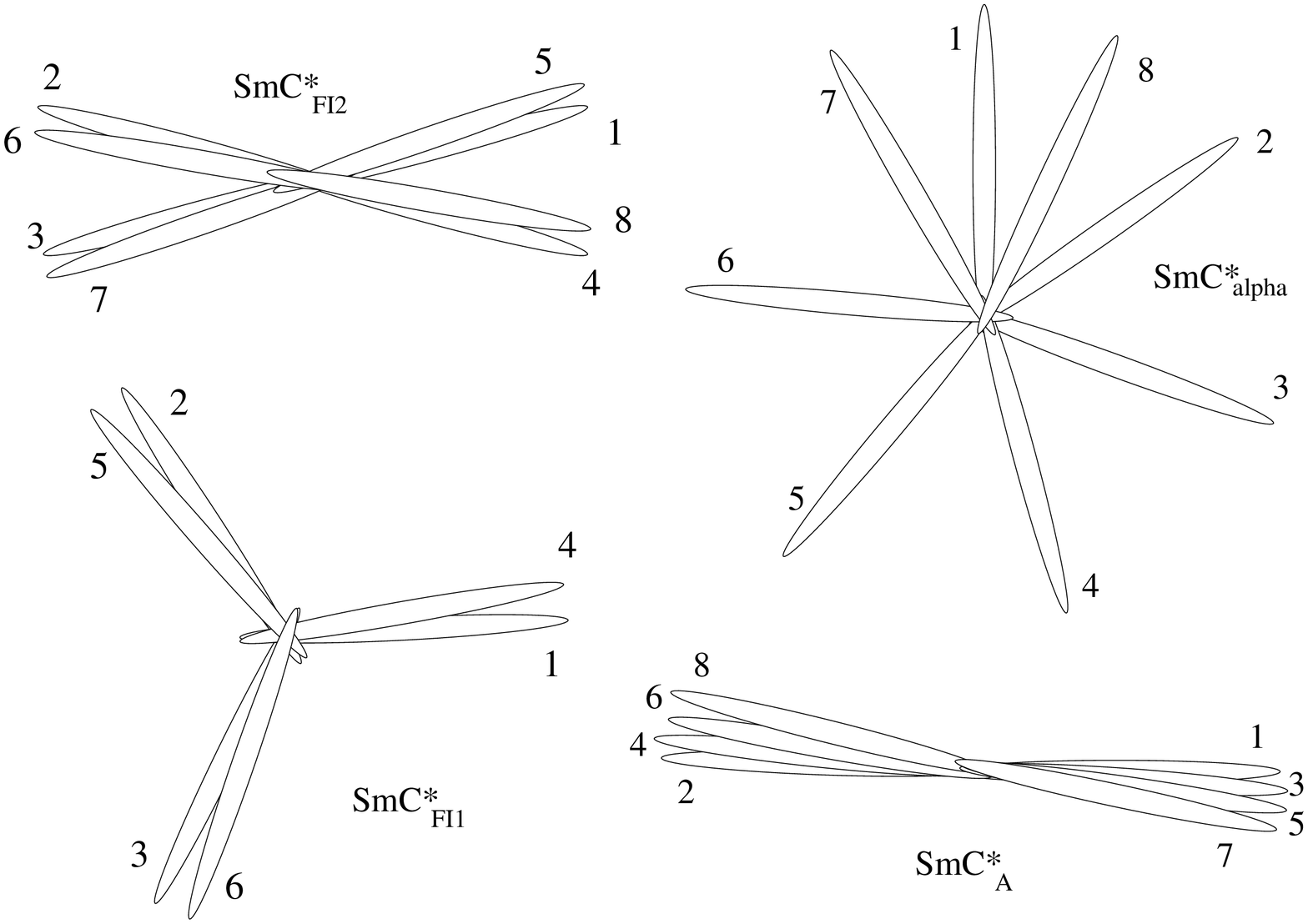}
\caption{Variation of the azimuthal angle of molecules for the
antiferroelectric subphases. The numbers indicate the plane sequence
(Adapted from
Hirst \textit{et al.}, 2002).}
\end{figure}

\subsection{Antiferroelectric phases}

Figure 3 shows the range of different antiferroelectric 
phases that can occur for chiral
materials with a SmC phase.
The most elaborate phase sequence for a chiral molecule studied using
resonant x-ray diffraction is for 10OTBBB1M7 which has the sequence
I--SmA--SmC$^*_{\alpha}$--SmC$^*$--SmC$^*_{FI2}$--SmC$^*_{FI1}$--SmC$^*_A$--K
on reducing the temperature (Mach \textit{et al.}, 1998; Mach
\textit{et al.}, 1999; Hirst \textit{et al.}, 2002). For materials with
fewer phases the order of those which remain is retained. The
SmC$^*_{FI1}$ and SmC$^*_{FI2}$ are three and four plane
antiferroelectric
phases. The resonant diffraction results have been compared to existing
theoretical models. Ising models had previously been used to explain the
antiferroelectric plane sequences (Fukuda \textit{et al.}, 1994). 
These required interlayer angles to be 0 or
$\pi$. By contrast a clock model allows for any constant angle
increment between planes (\v{C}epi\v{c} \& \v{Z}ek\v{s}, 1995; Lorman,
1995; Pikin \textit{et al.}, 1995; Roy \& Madushudana, 1996). 
Based on the experimental results more
detailed models are being developed.

As described above, for x-rays close in energy to
an absorption edge the scattering depends on the photon polarization
relative to the orientation of the 
molecule containing the electron. Determination of the
in-plane orientation of molecules in antiferroelectric 
phases becomes possible.
Mach \textit{et al.} (1998) presented the first
attempt in this direction and they showed clearly that different
antiferroelectric
structures had different stacking sequences in terms of azimuthal
orientation (see Fig.3). These studies involved chiral molecules
incorporating sulfur, which have an accessible K-edge. Subsequent studies
involved selenium (Matkin \textit{et al.}, 2001). 
In addition to the antiferroelectric stacking sequences
these materials have a helical pitch in the $\sim1$ $\mu$m range.
This research was a significant
advance for the field. The structure of SmC$^*_{\alpha}$ was here
determined for the first time (Mach \textit{et al.}, 1998).
These measurements were able to rule out Ising type models for the
antiferroelectric phases
in favour of a clock model. 

Mach \textit{et al.} (1999) 
have studied the SmC$^*_{\alpha}$, SmC$^*_{FI2}$, SmC$^*_{FI1}$,
SmC$^*_A$ phases in liquid crystals
containing sulfur while using polarization analysis to test the
predictions of the clock model. The resonant peaks were often observed
to be split (see Fig.4) due to the coupling between the short plane
sequence and the long helical pitch. In the clock model the first-order
satellites are predicted to be
$\mathbf{\pi}$ polarized. 
By contrast the
second-order reflections are 
$\mathbf{\sigma}$ polarized.
This was observed, confirming the polarization predictions
for the first time.

In spite of many years of research the
SmC$^*_{\alpha}$, SmC$^*_{FI2}$, SmC$^*_{FI1}$, SmC$^*_A$ phases are
still not fully understood.
Resonant x-ray scattering showed the stacking sequences for the first
time. Following studies of MHPOBC 
it was implied (Lagerwall \textit{et al.}, 2003)
that the intervention of the SmC$^*$ phase between SmC$^*_{\alpha}$ and 
SmC$^*_{FI2}$ phases could be due to sample degradation or impurity effects.
The SmC$^*_{\alpha}$ phase follows in temperature below the SmA phase.
In this phase the tilt angle is changing rapidly from zero as the
temperature is lowered. The plane sequence is 
incommensurate with the long helical pitch.
Hirst \textit{et al.} (2002) provided new
information concerning the SmC$^*_{\alpha}$ - SmC$^*$ - SmC$^*_{FI2}$
phase sequence with a SmC$^*$ phase which was stable for a long
temperature range. The periodicity varied from about 10 layers at higher
temperature
(SmC$^*_{\alpha}$) to about 75 layers at lower temperature from which
the peaks jump to the four layer repeat configuration known to
correspond to the SmC$^*_{FI2}$ structure. They observed a continuous
variation in the periodicity across the SmC$^*_{\alpha}$ - SmC$^*$
transition. The layer spacing is observed split above the
SmC$^*_{\alpha}$ - SmC$^*$ transition, an effect which was associated
with a sample degradation induced SmC$^*$ phase by Lagerwall \textit{et
al.} (2003) for a different antiferroelectric material. In the MHPOBC
case the layer spacing splitting occurred for all the 
tilted phases while for the new
material, studied with resonant techniques, the layer spacing 
splitting occurred mainly in
the SmC$^*_{\alpha}$ phase. Hirst \textit{et al.} (2002) suggest the
splitting is likely to be due to temperature gradients or surface
effects.

\begin{figure}[h]
\includegraphics[scale=0.65]{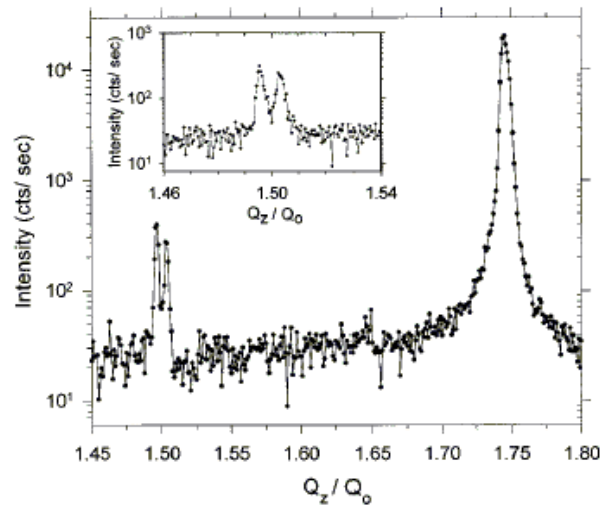}
\caption{X-ray intensity versus wave-vector transfer for the
SmC$^*_{FI2}$ phase of MHDDOPTCOB. (Reproduced with permission from
Mach \textit{et al.}, 1999.) The splitting becomes resolvable for the
second order peak.}
\end{figure}

The SmC$^*_{FI2}$ phase (Fig.3) was observed in 10OTBBB1M7 (Mach
\textit{et al.}, 1998) and in MHDDOPTCOB (Mach \textit{et al.}, 1999).
Lorman (1996) showed that the
SmC$^*_{FI2}$ phase could either involve a structure which is uniaxial
or biaxial about the layer normal. The former has a 90$^{\circ}$
rotation from one layer to the next while the latter rotates alternately
by $\delta$ and $180^{\circ} - \delta$. 
The early resonant x-ray results supported a uniaxial structure while
ellipsometry measurements (Johnson \textit{et al.}, 2000) showed that
the structure was biaxial. 
Cady 
\textit{et al.} (2001) confirmed using higher-resolution resonant x-ray
diffraction that it was indeed biaxial, with the
additional modification of constant interlayer rotation on 
$\sim1$ $\mu$m scale. The data and structure factors were in good agreement.
The combination of two modulations results in the splitting of resonant 
reflections shown in Fig.4.

The three plane repeat SmC$^*_{FI1}$ phase has been studied for several
materials (Mach \textit{et al.}, 1998; Mach \textit{et al.}, 1999; Hirst
\textit{et al.}, 2002). The three plane order and predicted polarization
have been confirmed (Mach \textit{et al.}, 1999). Fine structure was
observed in the resonant peak but this did not correspond to the
splitting due to the long helical pitch.

A wealth of information is now available to guide understanding of these
phases.
The Ising models for the SmC$^*$ phases, which suggest a single tilt
plane with combinations of 0
and $\pi$ rotation angles, are ruled out by
the X-ray results (Mach \textit{et al.}, 1998).
\v{C}epi\v{c} and \v{Z}ek\v{s} (2001) have created a phenomenological
model which is consistent with the observed SmC$^*_{FI2}$ and SmC$^*_{FI1}$
structures. The key observation is that the existence of subphases is
highly sensitive to the optical purity of the sample. This model
includes a variety of interactions: steric, van der Waals, 
electrostatic, piezoelectric,
flexoelectric between nearest layers and electrostatic between next 
nearest layers. The
cause and role of the discrete flexoelectric effect has been elaborated
by Emelyanenko and Osipov (2003). Using a similar model they predict the
observed phases and additionally subphases with longer plane sequences
which have not yet been observed.

Without establishing a specific model Lagerwall \textit{et al.} (2003)
suggest that the chirality is not the key ingredient. Using a range of
examples they point out that antiferroelectric subphase suppression
can be due to mixing induced steric and polar interactions with the role
of chirality being less important. Related observations are made: (1)
that the short plane stacking sequences and the SmC$^*_A$ phase are
incompatible with the interdigitation of layers which is common in
smectic materials. (2) that SmC$^*_A$ order is incompatible with the
existence of a high temperature N$^*$ phase; instead I-SmA transitions
occur. From this pair of observations it is concluded that the tendency
to form discrete layers is very strong in these materials. In addition
these phases are only observed for materials with large dipoles and
occur when the tilt angle is large. The suggestion is that electrostatic
and steric effects dominate with chirality playing a less important
role. A full model for the range of observed behaviour is not yet
available. That this approach could explain the SmC$^*_{FI2}$ and
SmC$^*_{FI1}$ structures is contested (Osipov, 2004).

\section{Smectic phases within gel structures}

In the studies reviewed in the 
previous two sections complicated phase transition behaviour was
observed in single component systems. In the studies reviewed in 
this section the novelty arises
due to the two different components: the liquid crystal and the gel
structure. The effect of disorder on the SmA state is fascinating
because
the state itself lacks true long-range order. Fluctuations in the layer
position diverge logarithmically with the size of the system.
This is known as the Landau-Peierls instability and was confirmed by 
Als-Nielsen
\textit{et al.} (1980). 
The results of SmA liquid crystals in gels 
are compared to bulk liquid crystal behaviour and
to theoretical predictions. The bulk phenomena of importance are the
coupling between the N and SmA phases and the anomalous elasticity
exhibited by
the SmA phase. The important theoretical predictions are those of the
random-field model (Imry \& Ma, 1975) 
and those due to Radzihovsky and Toner (1999). The bulk liquid crystal
features and the theoretical models are introduced in the following
paragraphs.

The N-SmA transition has unusual critical properties due to the coupling
between N and SmA order parameters and fluctuations. If the N-SmA
transition occurs close in temperature to the I-N transition, then the N
order is strongly enhanced by the SmA order (Garland \& Nounesis, 1994). 
This coupling leads to a
tricritical point (McMillan, 1971; de Gennes \& Prost, 1993). 
Conversely, if the N order is fully saturated before
the N-SmA transition, then the critical properties more closely resemble
those for a two-component order parameter (XY) system in three
dimensions (3D).
Coupling between the N fluctuations and the SmA order
parameter leads to an anisotropic critical regime. This picture has recently
been confirmed with unprecedented precision
via x-ray studies of samples aligned in a 5 T magnetic field (Primak 
\textit{et al.}, 2002a \& 2002b).
The theory of this
behaviour is not yet complete. Any changes to these couplings due to the
gel environment may well be instructive.

Anomalous elasticity in bulk SmA materials was first discussed by
Grinstein \& Pelcovits (1981). Displacements in the layering direction
and those perpendicular to the layers are coupled due to the non-linear
stress-strain relation (de Gennes \& Prost, 1993). Due to the
Landau-Peierls instability, there are always angular fluctuations. Hence
when the stress is zero the displacements in the in-plane direction
cause displacements in the layering direction. This  modifies the layer
thickness. An imposed stress can easily be accommodated by suppressing
the tilt fluctuations. A related effect occurs for the splay. 
The compression, B, and splay, K$_1$, elastic
constants vanish and diverge respectively on long length scales.
Disorder may modify the tilt fluctuations and this may be reflected in
the elastic properties.

Two models for the effect of the gel are considered. The first is the
random-field model (Larkin \& Ovchinnikov, 1979) corresponding to the
pinning of the layer position by the random disorder.
The N-SmA transition is described by a
two component order parameter $\psi(\mathbf{r}) = |\psi(\mathbf{r})|
e^{i q_0 u(\mathbf{r})}$ where $|\psi(\mathbf{r})|$ is the strength of
the density modulation, $q_0 = 2\pi / d$ where $d$ is the layer
periodicity, and $u(\mathbf{r})$ is the layer displacements away from
perfect order.  
The contribution to the smectic free
energy density is
\begin{equation}
f_{rf} = -Re(V(\mathbf{r}) \psi(\mathbf{r})).
\end{equation}

\noindent The random disorder, $V(\mathbf{r})$, pins the phase of the density
wave. These random fields would be expected to
destroy the ordered state, replacing it with finite sized domains (Imry
\& Ma, 1975). The size of the domains and the structure factor are
predicted to vary systematically with the variance of the layer pinning
field (Aharony \& Pytte, 1983). The changes depend on the lower marginal
dimensionality of the system.

The second model is due to Radzihovsky and Toner
(1999). They
emphasise two important effects of the gel on the liquid crystal: the
pinning of the position of the smectic layers and the pinning of the
orientation of the nematic director. The contribution to the smectic
free energy density is
\begin{equation}
f_{gel} = f_{rf} -(\mathbf{g}(\mathbf{r})\mathbf{.\hat{n}}(\mathbf{r}))^2. 
\end{equation}

\noindent In this case 
$\mathbf{g}$($\mathbf{r}$) represents a random tilt effect imposed by
the gel on the director.
The effect of the field, $V(\mathbf{r})$, 
is to destroy the ordered phase, replacing it with finite
sized ordered domains since this is the random-field model described
above. The effect of the field, $\mathbf{g}(\mathbf{r})$, which
pins the orientation of the nematic director, is to perturb the smectic
order via the interaction between the two order parameters and their
fluctuations.
Radzihovsky and Toner (1999) found that the effect of disorder on the
director would produce novel changes on the smectic state. A new smectic
Bragg glass state was predicted characterized by short-range smectic
order, glassy dynamics and enhanced anomalous elasticity.
Fluctuations
created by the perturbation of the director strongly enhance
the anomalous elasticity effect (Radzihovsky \& Toner, 1999). 
This model predicts that the correlation length should have a
temperature dependence related to that of the layer compression modulus, B
(Bellini \textit{et al.}, 2001).

In the studies reviewed
here liquid crystals are confined in chemically bonded porous silica
structures (aerogels) or alternatively in a weakly bonded gel formed by
hydrogen bonds between silica
nanoparticles (aerosils).
With increasing quantities of silica the samples became progressively
more opaque to x-rays.
The availability of high brilliance synchrotron x-ray sources has
facilitated the study of liquid crystal phase transitions within a
silica gel environment.

\begin{figure}[h]
\includegraphics[scale=0.5]{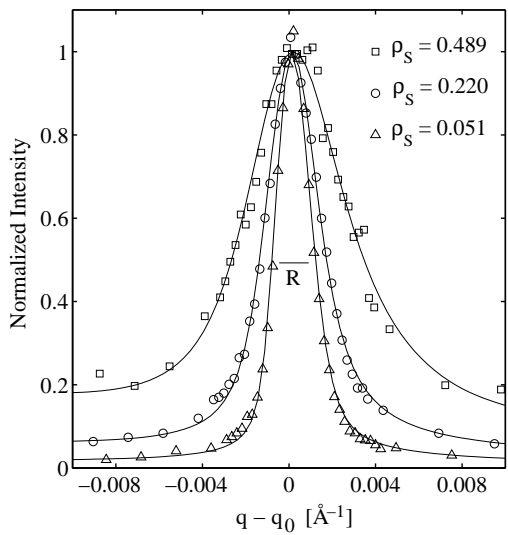}
\caption{The broadening of the normalized x-ray reflection from
SmA order due to the influence of the aerosil gel structure. In this
case the results are for 8OCB-aerosil at low temperatures (approximately
$T^0_{NA} - 25$ K). The horizontal line labeled R is the
width of the instrumental resolution. The model of the effect of the gel
is shown by the lines through the points.
(Reproduced from Clegg,
\textit{et al.}, 2003a.)}
\end{figure}

\subsection{N-SmA transition in gels}

The N-SmA transition of 8CB was studied using x-ray
diffraction within a rigid silica aerogel (Clark \textit{et al.}, 1993; 
Rappaport \textit{et al.},
1996; Bellini \textit{et al.}, 2001). The N-SmA transition was
observed to be destroyed as seen by the broadening of the corresponding
x-ray reflection.
Comparisons of
the temperature dependence of the correlation length and peak amplitude
suggested that the 8CB-aerogel system is close to the novel smectic Bragg
glass state (Bellini \textit{et al.}, 2001). It was subsequently a surprise to find that weakening the
disorder by changing from aerogel to aerosil did not move the system
closer to this novel soft state.

High-resolution x-ray diffraction studies were carried-out on
8CB-aerosil gels (Park \textit{et al.}, 2002; Leheny \textit{et al.},
2003) and 8OCB-aerosil gels (Clegg \textit{et al.}, 2003a). The gel is
fabricated from aerosils which are hydrophilic silica nanoparticles.
These are dispersed in the liquid crystal and they hydrogen bond
together to create a very low density gel structure. The gel percolates
above a very low threshold volume fraction and the liquid crystal gel
forms a soft composite.
A range of gel densities were explored over the temperature range
of the N-SmA transition.
The smectic
correlation length was found to saturate at a finite value at low
temperatures and this value depended on the density of the gel
structure (Fig.5). 
Both the 3D-XY random-field model and the model due to Radzihovsky and
Toner (1999)
predict that the structure factor characterizing the smectic
fluctuations will be altered in the gel.
The solid lines in Fig.5 are the result of fits of the anticipated
two-component line-shape to 8OCB-aerosil data.
The finite correlation length indicates that the QLRO SmA
phase has been destroyed, although there remains a pseudo-transition
between a high temperature regime where thermal fluctuations dominate
and a low temperature regime where static quenched-random effects
dominate.
A further study was
carried out on the highly anisotropic liquid crystal $\overline{8}S5$ in
an aerosil gel (Clegg \textit{et al.}, 2003b). The two-component
line-shape also gave a very good account of these results. A low
temperature correlation length could not be extracted in this case due
to the intervention of the SmA-SmC phase transition.

The correlation lengths at low temperature for 8CB-aerosil and
8OCB-aerosil are plotted in Fig.6 as a function of the density of the
gel ($\rho_S$ = mass of silica / volume of liquid crystal). 
Since each new particle perturbs the average phase which is
favourable to the existing aerosils,
the gel density is taken to be a measurement of the variance
of the random field. The solid lines in Fig.6 are the expected
behaviour according to 3D-XY random-field model (Aharony \& Pytte, 1983). 
The good
agreement suggests that layer pinning may be the dominant effect in these
systems.

Comparisons to the random-field model have been augmented 
by more detailed observations. Systematic changes in the pseudo-critical
properties have been recorded in all cases. This is most evident using
ac-calorimetry (Zhou \textit{et al.}, 1997;
Iannacchione \textit{et al.}, 1998;
Clegg \textit{et
al.}, 2003a). The heat capacity 
exponent, $\alpha$, reflects the N susceptibility at the N-SmA
transition. High N susceptibility gives an $\alpha$ close to the
tricritical value. Low N susceptibility implies 
little coupling between N and SmA order parameters and 
$\alpha$ approaches the 3D-XY value. For the N-SmA transition in an
aerosil gel $\alpha$ moves
toward the 3D-XY value with increasing gel density. Comparable behaviour
is observed for the disconnected susceptibility as studied using x-ray
diffraction (Leheny \textit{et al.}, 2003; 
Iannacchione \textit{et al.}, 2003).
The evidence suggests that the N-SmA anisotropic criticality moves
toward 3D-XY pseudo-criticality 
in the presence of quenched random disorder. This
observation is consistent with the gel causing the N susceptibility to
decrease toward zero. This effect is not currently part of either
theoretical model.

One of the key differences between x-ray results for the N-SmA in
aerogel and aerosil is the temperature dependence of the correlation
lengths and the susceptibility. Both show these quantities saturating at
finite values at low temperature. In aerogel the temperature dependence
in the transition region
is strongly altered from bulk behaviour and this has been interpreted to
suggest enhanced anomalous elasticity (Bellini \textit{et al.}, 2001).
By contrast, with aerosil the temperature dependence in the transition
region is only subtly
modified from bulk behaviour (Leheny \textit{et al.}, 2003; 
Iannacchione \textit{et al.}, 2003). The
modification correlates well with ac-calorimetry results.

\begin{figure}[h]
\includegraphics[scale=0.5]{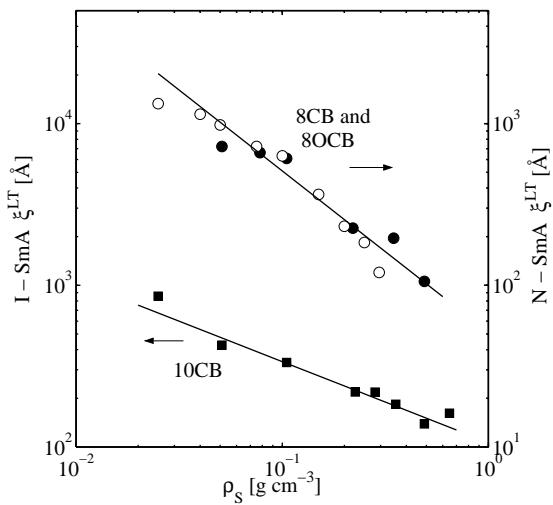}
\caption{Cube root of the SmA correlation volume, $\overline{\xi}$,
taken at low temperature versus the gel density,
$\rho_S$, for N-SmA and I-SmA transitions. The solid lines represent single
power laws which are different for the two types of transition.
The open circles are for 8CB-aerosil (Leheny \textit{et al.}, 2003), the
solid circles are for 8OCB-aerosil (Clegg, \textit{et al.}, 2003a) and
the solid
squares are for 10CB-aerosil (Ramazanoglu \textit{et al.}, 2004).
(Reproduced from Ramazanoglu \textit{et al.}, 2004.)}
\end{figure}

\subsection{The role of nematic order}

The apparent schism between aerogel and aerosil results
motivated further studies particularly
concerning the role of nematic order.
Ramazanoglu \textit{et al.} (2004) studied the I - SmA transition in
10CB-aerosil where there is no orientational order prior to the
formation of SmA order.
By contrast Liang \textit{et al.} (2004) studied the N-SmA
transition in an anisotropic aerosil gel.
In this case the liquid
crystal has long-range orientational order prior to the onset of SmA
order. These studies will be described in turn.

In order to probe the interaction between the orientational order and
the gel environment Ramazanoglu \textit{et al.} (2004) 
studied 10CB-aerosil gels.
Bulk 10CB has a strongly
first-order transition directly from an I phase into the
SmA phase. In the aerosil gel the SmA correlation length, extracted
using the two component line shape model, was found to saturate at low
temperatures and to decrease systematically with increasing gel density.
As shown in Fig.6, the density dependence was $1/\surd{\rho_S}$ for the
I-SmA transition while it was $1/\rho_S$ for the N-SmA transition. This
difference, resulting from the presence of nematic order, is consistent
with the random-field model.
The change is associated with the fact that the nematic director is forced
to point in the direction of the layer normal
(Ramazanoglu \textit{et al.}, 2004). Preliminary measurements were made
on the I-SmA transitions for
10CB and 650BC in rigid aerogel (Bellini \textit{et al.}, 1996; Bellini
\textit{et al.}, 2003). The
correlation lengths are observed to increase discontinuously at the
transition temperature and to saturate quickly. The scattering intensity
increases only slowly with decreasing temperature. The results appear
very similar to those for aerosil gels at high $\rho_S$. Only a single
aerogel pore size was studied making a full comparison with the aerosil
results difficult at present.

\begin{figure}
\includegraphics[scale=0.7]{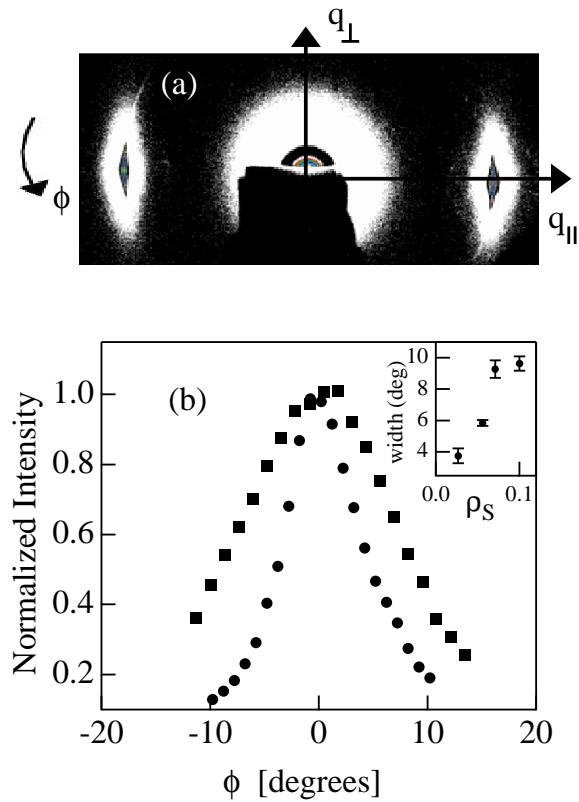}
\caption{(a) X-ray intensity distribution from SmA correlations in an
anisotropic aerosil gel ($\rho_S = 0.035$ g cm$^{-3}$). The underlying
nematic order is long range while the SmA order is short range. (b)
The sample mosaic spread for anisotropic gel samples of density $\rho_S
= 0.027$ g cm$^{-3}$ (circles) and $\rho_S = 0.100$ g cm$^{-3}$ (squares) at
300.1 K. (Reproduced with permission from Liang \textit{et al.},
2004.) }
\end{figure}

Liang \textit{et al.} (2004) 
created anisotropic gels by \textit{training} standard liquid
crystal aerosil gels in a magnetic field. The idea for this method
followed from the deuterium NMR studies by Jin and Finotello (2001).
The alignment of the molecules by the field led to the gradual but
permanent rearrangement of the gel structure. This technique resulted in
anisotropic gels provided the aerosil density $\rho_S \leq 0.1$ g
cm$^{-3}$.
High-resolution x-ray studies were carried out on 8CB in an anisotropic
gel across the temperature range of the N-SmA transition. In part, this
research was to test a theoretical prediction
that an XY Bragg glass phase will be stabilized in smectics with
anisotropic disorder (Jacobsen \textit{et al.}, 1999). 
This prediction is for the case of disorder which
has been stretched so that the nematic director has a preferred
orientation. This situation suppresses long length-scale nematic
fluctuations.
The low temperature
state should then be an XY Bragg glass state with algebraically decaying
correlations.

The studies showed that long-range nematic
order was created by training the gels as seen via the anisotropy in Fig. 7. 
The SmA correlations which developed below
the pseudo-transition were not characteristic of the anticipated XY 
Bragg
glass state (Liang \textit{et al}., 2004). 
The behaviour through and below the pseudo-transition has
many similarities with the behaviour of 8CB-aerosil when the gel is
isotropic. In the trained case the signal to noise ratio is highly
favourable since the sample is a \textit{single crystal}. The structure
factor was found to be modified slightly compared to the model
for the untrained system.
The anisotropic scaling of correlation lengths is suppressed
when the random-fields are imposed.
This is indicative of the anisotropic N-SmA
fixed point becoming more like 3D-XY behaviour in the presence of
quenched disorder. It appears that both couplings between N and SmA
order parameters and fluctuations are suppressed by the random environment. 

The series of
measurements on the N-SmA transition in aerosil gels have failed to
corroborate earlier indications of smectic Bragg glass behaviour in an
aerogel. In addition, both N-SmA and I-SmA transitions are modified in a
manner consistent with a simple random-field model. It may be possible
that the earlier conclusions with respect to the 
smectic Bragg glass 
(Bellini \textit{et al.}, 2001) were premature.

\section{Conclusions}

This review has shown the huge range of behaviour exhibited by smectic
liquid crystals when they are perturbed by twist, dipole moments, and
local deformations. This has included a rich
variety of phases and novel pseudotransition behaviour. Many of the
observations are still incompletely understood.

Further developments using x-ray techniques are anticipated in the three
areas described.
In the field of twist-grain-boundary studies
recent observations reveal further new phases, these have not yet
been subjected to x-ray study in oriented samples and have been reviewed
by Goodby (2002). The range includes TGBC$^*_A$ structures 
where the SmC$^*_A$
order is antiferroelectric (Goodby \textit{et al}, 2000) 
and also reentrant TGBA phases 
(Rao \textit{et al.}, 2001).

Resonant x-ray diffraction can now be applied to a much broader range of
materials.
One key advance was the demonstration that
non-resonant mesogens could be studied using resonant x-ray diffraction
by mixing them with a small fraction of resonant liquid crystal (Hirst
\textit{et al.}, 2002). The second advance was the application of
resonant diffraction to the study of liquid crystals containing Se
(Matkins \textit{et al.}, 2001). The energy of the K-edge (12.66 keV) is
sufficient to allow samples to be studied in the context of realistic
electro-optic devices.

The stability of the smectic Bragg glass 
could be further probed by combining the
8CB-aerogel samples studied by Bellini \textit{et al} (2001) with a high
magnetic field. 
The effects initially observed
should be enhanced as the N order becomes long-range. Orienting the 
sample would also improve the signal to noise ratio without increasing
the sample damage. Recent pioneering
research 
includes new studies on SmC$^*$ phases in an aerosil gel
(Kutnjak \textit{et al.}, 2003) which have yet to be examined with x-ray
diffraction.

\section{Acknowledgements}

I am grateful to S.Egelhaaf, 
C.Garland, H.Gleeson, G.Iannacchione, R.Leheny,
M.Osipov and R.Sastri for helpful comments and to
P.Barois, H.Gleeson, R.Leheny, L.Navailles and R.Pindak
for providing figures.
Funding in Edinburgh came from the EPSRC (Grant GR/S10377/01). 
Much of the research presented 
in Section 4 was carried out at the University of Toronto and was funded
by the Natural Sciences and Engineering Research Council of Canada.


\begin{itemize}
\item[ ] Abrikosov, A.A. (1957). \emph{Sov. Phys. JETP} \textbf{5},
1174--1183.
\item[ ] Aharony, A. \& Pytte, E. (1983). \emph{Phys. Rev. B}
\textbf{27}, 5872--5874.
\item[ ] Als-Nielsen, J., Litster, J.D., Birgeneau, R.J., Kaplan, M.,
Safinya, C.R., Lindegaard-Andersen, A. \& Mathiesen, S. (1980).
\emph{Phys. Rev. B} \textbf{22}, 312--320.
\item[ ] Als-Nielsen, J. \& McMorrow, D. (2001). \emph{Elements of
Modern X-ray Physics} Chichester: John Wiley.
\item[ ] Barois, P., Heidelbach, F., Navailles, L., Nguyen, H.T.,
Nobili, M., Petit, M., Pindak, R. \& Riekel, C. (1999). \emph{Eur. Phys.
J. B} \textbf{11}, 455--462.
\item[ ] Bellini, T., Rappaport, A.G., Clark, N.A. \& Thomas, B.N
(1996). \emph{Phys. Rev. Lett.} \textbf{77}, 2507--2510.
\item[ ] Bellini, T., Radzihovsky, L., Toner, J. \& Clark, N.A.
(2001). \emph{Nature} \textbf{294}, 1074--1079.
\item[ ] Bellini, T., Clark, N.A. \& Link, D.R. (2003). \emph{J.
Phys.: Condens. Matter} \textbf{15}, S175--S182.
\item[ ] Brunet, M., Navailles, L. \& Clark, N.A. (2002). \emph{Eur.
Phys. J. E} \textbf{7}, 5--11.
\item[ ] Cady, A., Pitney, J.A., Pindak, R., Matkin, L.S., Watson,
S.J., Gleeson, H.F., Cluzeau, P., Barois, P., Levelut, A.-M., Caliebe,
W., Goodby, J.W., Hird, M. \& Huang, C.C. (2001). \emph{Phys. Rev. E}
\textbf{64}, 050702(R)-1--4.
\item[ ] \v{C}epi\v{c}, M. \& \v{Z}ek\v{s}, B. (1995). \emph{Mol.
Cryst. Liq. Cryst. A} \textbf{263}, 61--67.
\item[ ] \v{C}epi\v{c}, M. \& \v{Z}ek\v{s}, B. (2001). \emph{Phys.
Rev. Lett.} \textbf{87}, 085501-1--4.
\item[ ] Chan, T. \& Garland, C.W. (1995). \emph{Phys. Rev. E}
\textbf{52}, 5000--5003.
\item[ ] Clark, N.A., Bellini, T., Malzbender, R.M., Thomas, B.N.,
Rappaport, A.G., Muzny, C.D., Schaefer, D.W. \& Hrubesh, L. (1993). 
\emph{Phys. Rev. Lett.} \textbf{71},
3505--3508.
\item[ ] Clegg, P.S., Stock, C., Birgeneau, R.J., Garland, C.W., Roshi, A. 
\& Iannacchione, G.S. (2003a). \emph{Phys. Rev. E} 
\textbf{67}, 021703-1--13.
\item[ ] Clegg, P.S., Birgeneau, R.J., Park, S., Garland, C.W., 
Iannacchione, G.S., Leheny, R.L. \& Neubert, M.E. (2003b). \emph{Phys. Rev. E} 
\textbf{68}, 031706-1--7.
\item[ ] Davidov, D., Safinya, C.R., Kaplan, M., Dana, S.S.,
Schaetzing, R., Birgeneau, R.J. \& Litster, J.D. (1979). \emph{Phys.
Rev. B} \textbf{19}, 1657--1663.
\item[ ] de Gennes, P.G. (1972). \emph{Solid State Commun.} \textbf{10}, 
753--756.
\item[ ] de Gennes, P.G. \& Prost, J. (1993). \emph{The Physics of
Liquid Crystals}. 2nd edition: New York: Oxford University Press.
\item[ ] Dmitrienko, V.E. (1983). \emph{Acta Cryst.} A\textbf{39},
29--35.
\item[ ] Dozov, I. (1995). \emph{Phys. Rev. Lett.} \textbf{74},
4245--4248.
\item[ ] Emelyanenko, A.V. \& Osipov, M.A. (2003). \emph{Phys. Rev. E}
\textbf{68}, 051703-1--15.
\item[ ] Fera, A., Opitz, R., de Jeu, W.H., Ostrovskii, B.I., Schlauf,
D. \& Bahr, Ch. (2001). \emph{Phys. Rev. E} \textbf{64}, 021702-1--8.
\item[ ] Fukuda, A., Takanishi, Y., Isozaki, T., Ishikawa, K. \&
Takezoe, H. (1994). \emph{J. Mater. Chem.} \textbf{4}, 997--1016.
\item[ ] Furukawa, K., Tereshami, K., Ichihashi, M., Saitoh, S.,
Miyazawa, K. \& Inukai, T. (1988). \emph{Ferroelectrics} \textbf{85},
451--451.
\item[ ] Galerne, Y. (2000). \emph{Eur. Phys. J. E} \textbf{3},
355--368.
\item[ ] Garland, C.W. \& Nounesis, G. (1994). \emph{Phys. Rev. E}
\textbf{49}, 2964--2971.
\item[ ] Goodby, J.W., Waugh, M.A., Stein, S.M., Chin, E., Pindak, R.
\& Patel, J.S. (1989a). \emph{Nature} \textbf{337}, 449--452.
\item[ ] Goodby, J.W., Waugh, M.A., Stein, S.M., Chin, E., Pindak, R.
\& Patel, J.S. (1989b). \emph{Am. Chem. Soc.} \textbf{111}, 8119--8125.
\item[ ] Goodby, J.W., Petrenko, A., Hird, M., Lewis, R.A., Meier, J. 
\& Jones, J.C. (2000). \emph{J. Chem. Soc. Chem. Commun.}, 1149--1150.
\item[ ] Goodby, J.W. (2002). \emph{Curr. Opin. Colloid In.} \textbf{7}, 
326--332.
\item[ ] Grinstein, G. \& Pelcovits, R.A. (1981). \emph{Phys. Rev.
Lett.} \textbf{47}, 856--859.
\item[ ] Hiji, N., Chandani, A.D.L., Nishiyama, S., Ouchi, Y.,
Takezoe, H. \& Fukuda, A. (1988). \emph{Ferroelectrics} \textbf{85},
99--99.
\item[ ] Hirst, L.S., Watson, S.J., Gleeson, H.F., Cluzeau, P.,
Barois, P., Pindak, R., Pitney, J., Cady, A., Johnson, P.M., Huang,
C.C., Levelut, A.-M., Srajer, G., Pollmann, J., Caliebe, W., Seed, A.,
Herbert, M.R., Goodby, J.W. \& Hird, M. (2002). \emph{Phys. Rev. E}
\textbf{65}, 041705-1--10.
\item[ ] Iannacchione, G.S., Garland, C.W., Mang, J.T. \& Rieker, T.P. 
(1998). \emph{Phys. Rev. E} \textbf{58}, 5966--5981.
\item[ ] Iannacchione, G.S., Park, S., Garland, C.W., Birgeneau, R.J.
\& Leheny, R. (2003). \emph{Phys. Rev. E} \textbf{67} 011709-1--13
\item[ ] Imry, Y. \& Ma, S.-K. (1975). \emph{Phys. Rev. Lett.}
\textbf{35}, 1399--1401.
\item[ ] Jacobsen, B., Saunders, K., Radzihovsky, L. \& Toner, J.
(1999). \emph{Phys. Rev. Lett.} \textbf{83}, 1363--1366.
\item[ ] Jin, T. \& Finotello, D. (2001). \emph{Phys. Rev. Lett.}
\textbf{86}, 818--821.
\item[ ] Johnson, P.M., Olson, D.A., Pankratz, S., Nguyen, T., Goodby,
J., Hird, M. \& Huang, C.C. (2000). \emph{Phys. Rev. Lett.} \textbf{84},
4870--4873.
\item[ ] Kamien, R.D. \& Lubensky, T.C. (1993). \emph{J. Phys. I
(France)} \textbf{3}, 2131--2138.
\item[ ] Kumar, S. (2001). \emph{Liquid Crystals, Experimental study
of physical properties and phase transitions}. edited Kumar, S.:
Cambridge: Cambridge University Press.
\item[ ] Kutnjak, Z., Cordoyiannis, G. \& Nounesis, G. (2003).
\emph{Ferroelectrics} \textbf{294}, 105--111.
\item[ ] Lagerwall, J.P.F., Rudquist, P., Lagerwall, S.T. \&
Gie\ss elmann, F. (2003). \emph{Liq. Cryst.} \textbf{30}, 399--414.
\item[ ] Larkin A.I. \& Ovchinnikov, Yu.N. (1979). \emph{J. Low Temp.
Phys.} \textbf{34}, 409--428.
\item[ ] Leheny, R.L., Park, S., Birgeneau, R.J., Gallani, J.-L.,
Garland, C.W. \& Iannacchione, G.S. (2003). \emph{Phys. Rev. E} \textbf{67}, 
011708-1--13.
\item[ ] Levelut, A.-M. \& Pansu, B. (1999). \emph{Phys. Rev. E}
\textbf{60}, 6803--6815.
\item[ ] Liang, D., Borthwick, M.A. \& Leheny, R.L. (2004).
\emph{J.Phys.: Condens. Matter}
\textbf{16}, S1989--S2002.
\item[ ] Lorman, V.L. (1995). \emph{Mol. Cryst. Liq. Cryst. A} \textbf{262}, 437--453.
\item[ ] Lorman, V.L. (1996). \emph{Liq. Cryst.} \textbf{20},
267--276.
\item[ ] Mach, P., Pindak, R., Levelut, A.-M., Barois, P., Nguyen,
H.T., Huang, C.C. \& Furenlid, L. (1998). \emph{Phys. Rev. Lett.}
\textbf{81}, 1015--1018.
\item[ ] Mach, P., Pindak, R., Levelut, A.-M., Barois, P., Nguyen,
H.T., Baltes, H., Hird, M., Toyne, K., Seed, A., Goodby, J.W., Huang,
C.C. \& Furenlid, L. (1999). \emph{Phys. Rev. E} \textbf{60},
6793--6802.
\item[ ] Matkin, L.S., Watson, S.J., Gleeson, H.F., Pindak, R.,
Pitney, J., Johnson, P.M., Huang, C.C., Barois, P., Levelut, A.-M.,
Srajer, G., Pollmann, J., Goodby, J.W. \& Hird, M. (2001). \emph{Phys.
Rev. E} \textbf{64}, 021705-1--6.
\item[ ] McMillan, W.L. (1971). \emph{Phys. Rev. A} \textbf{4}, 1238--1246.
\item[ ] Meyer, R.B., Liebert, L., Strzelecki, L. \& Keller, P. (1975).
\emph{J. Phys. (Paris) Lett.} \textbf{36}, L69--L71.
\item[ ] Navailles, L., Barois, P. \& Nguyen, H.T. (1993). \emph{Phys.
Rev. Lett.} \textbf{71}, 545--548.
\item[ ] Navailles, L., Pindak, R., Barois, P. \& Nguyen, H.T. (1995).
\emph{Phys. Rev. Lett.} \textbf{74}, 5224--5227.
\item[ ] Navailles, L., Pansu, B., Gorre-Talini, L. \& Nguyen, H.T.
(1998). \emph{Phys. Rev. Lett.} \textbf{81}, 4168--4171.
\item[ ] Nguyen, H.T., Bouchta, A., Navailles, L., Barois, P., Isaert,
N., Tweig, R.J., Maaroufi, A. \& Destrade, C. (1992). \emph{J. Phys. II
France} \textbf{2}, 1889--1906.
\item[ ] Osipov, M.A. (2004). \emph{Private Communication}.
\item[ ] Park, S., Leheny, R.L., Birgeneau, R.J., Gallani, J.-L.,
Garland, C.W. \& Iannacchione, G.S. (2002). \emph{Phys. Rev. E}
\textbf{65}, 050703(R)-1--4.
\item[ ] Pershan, P.S. (1988). \emph{Structure of Liquid Crystal
Phases}. Singapore: World Scientific.
\item[ ] Petit, M., Barois, P. \& Nguyen, H.T. (1996). \emph{Europhys. Lett.}
\textbf{36}, 185--190.
\item[ ] Pikin, S.A., Hiller, S. \& Haase, W. (1995). \emph{Mol.
Cryst. Liq. Cryst. A} \textbf{262}, 425--435.
\item[ ] Pramod, P.A., Pratibha, R. \& Madhusudana, N.V. (1997).
\emph{Current Sci.} \textbf{73}, 761--765.
\item[ ] Primak, A., Fisch, M. \& Kumar, S. (2002a). \emph{Phys. Rev.
Lett.}
\textbf{88}, 035701-1--4.
\item[ ] Primak, A., Fisch, M. \& Kumar, S. (2002b). \emph{Phys. Rev. E}
\textbf{66}, 051707-1--13.
\item[ ] Radzihovsky, L. \& Toner, J. (1999). \emph{Phys. Rev. B}
\textbf{60}, 206--257.
\item[ ] Ramazanoglu, M.K., Clegg, P.S., Birgeneau, R.J., Garland, C.W.,
Neubert, M.E. \& Kim, J.M. (2004). \emph{Phys. Rev. E} \textbf{69}, 
061706-1--8.
\item[ ] Rao, D.S.S., Prasad, S.K., Raja, V.N., Yelemaggad, C.V. \&
Nagamani, S.A. (2001). \emph{Phys. Rev. Lett.} \textbf{87}, 085504-1--4.
\item[ ] Rappaport, A.G., Clark, N.A., Thomas, B.N. \& Bellini, T.
(1996). \emph{Liquid Crystals in Complex Geometries Formed by Polymer and
Porous Networks}. edited by Crawford, G.P. \& Zumer, S.: London: Taylor
and Francis.
\item[ ] Roy, A. \& Madushudana, N. (1996). \emph{Europhys. Lett.}
\textbf{36}, 221--226.
\item[ ] Renn, S.R. \& Lubensky, T.C. (1988). \emph{Phys. Rev. A}
\textbf{38}, 2132--2147.
\item[ ] Renn, S.R. \& Lubensky, T.C. (1991). \emph{Mol. Cryst. Liq.
Cryst.} \textbf{209}, 349--355. 
\item[ ] Renn, S.R. (1992). \emph{Phys. Rev. A} \textbf{45}, 953--973.
\item[ ] Ribeiro, A.C., Barois, Ph., Galerne, Y., Oswold, L. \&
Guillon, D. (1999). \emph{Eur. Phys. J. B} \textbf{11}, 121-126.
\item[ ] Safinya, C.R., Kaplan, M., Als-Nielsen, J., Birgeneau, R.J.,
Davidov, D., Litster, J.D., Johnson, D.L. \& Neubert, M.E. (1980). 
\emph{Phys. Rev. B} \textbf{21}, 
4149--4153.
\item[ ] Shashidhar, R., Ratna, B.R., Nair, G.G., Prasad, S.K., Bahr,
Ch. \& Heppke, G. (1988). \emph{Phys. Rev. Lett.} \textbf{61}, 547--549.
\item[ ] Seddon, J.M. (1998). \emph{Handbook of Liquid Crystals}. Vol.
1: edited Demus, D., Goodby, J., Gray, G.W., Spiess, H.-W. \& Vill, V.:
Weinheim: Wiley-VCH.
\item[ ] Takanishi, Y., Ikeda, A., Takezoe, H. \& Fukuda, A. (1995).
\emph{Phys. Rev. E} \textbf{51} 400--406.
\item[ ] Templeton, D.H. \& Templeton, L.K (1982). \emph{Acta Cryst.}
A\textbf{38}, 62--67.
\item[ ] Ybert, C., Navailles, L., Pansu, B., Rieutord, F., Nguyen,
H.T. \& Barois, P. (2003). \emph{Europhys. Lett.} \textbf{63}, 840--845.
\item[ ] Zhou, B., Iannacchione, G.S. \& Garland, C.W. (1997). 
\emph{Phys. Rev. E} \textbf{55}, 2962--2968.
\end{itemize}

\end{document}